\documentclass[structabstract]{aa}

\usepackage{amsmath}
\usepackage{graphicx}
\usepackage{txfonts}
\usepackage{float}

\begin{document}

\title{The jets of AGN as giant coaxial cables.}

\author{
Denise C. Gabuzda,\inst{1}
Matt Nagle\inst{1}
\and
Naomi Roche\inst{1} }

\institute{
Dept. of Physics, University College Cork, Cork, Ireland. \\
                 Email: d.gabuzda@ucc.ie\\
}

\def\gsim{\mathrel{\raise.5ex\hbox{$>$}\mkern-14mu
             \lower0.6ex\hbox{$\sim$}}}

\def\lsim{\mathrel{\raise.3ex\hbox{$<$}\mkern-14mu
             \lower0.6ex\hbox{$\sim$}}}

\date {Received ; accepted }

\abstract
{The currents carried by the jets of active galactic nuclei (AGNs)
can be probed using maps of the Faraday rotation measure (RM), since
a jet current will be accompanied by a toroidal magnetic
field, which will give rise to a systematic change in the
RM across the jet.}
{The aim of this study is to identify new AGNs displaying statistically
significant transverse RM gradients across their parsec-scale jets, in 
order to determine how often helical magnetic fields occur in AGN jets,
and to look for overall patterns in the implied directions for the
toroidal field components and jet currents.  }
{We have carried out new analyses of Faraday RM maps derived from previously
published 8.1, 8.4, 12.1 and 15.3~GHz data obtained in 2006 on the NRAO Very 
Long Baseline Array (VLBA). In a number of key ways, our procedures were
identical to those of the original authors, but the new imaging and 
analysis differs from the original methods in several ways: the technique used to match
the resolutions at the different frequencies, limits on the widths spanned
by the RM gradients analyzed, treatment of core-region RM gradients, 
approach to estimation of the significances of the gradients analyzed, and 
inclusion of a supplementary analysis using circular beams with areas equal 
to those of the corresponding elliptical naturally weighted beams. 
}
{This new analysis has substantially increased the number of AGNs known to
display transverse RM gradients that may reflect the presence of 
a toroidal magnetic-field component. The
collected data on parsec and kiloparsec scales indicate that the current
typically flows inward along the jet axis and outward in a more extended
region surrounding the jet, typical to the current structure of a co-axial
cable, accompanied by a self-consistent system of
nested helical magnetic fields, whose toroidal components give rise to
the observed transverse Faraday rotation gradients.  }
{The new results
presented here make it possible for the first time to conclusively
demonstrate the existence of a preferred direction for the toroidal
magnetic-field components --- and therefore of the currents --- of AGN jets.
Discerning the origin
of this current--field system is of cardinal importance for understanding
the physical mechanisms leading to the formation of the intrinsic jet
magnetic field, which likely plays an important role in the propagation and
collimation of the jets; one possibility is the action of a ``cosmic battery''.
}


\authorrunning{Gabuzda et al. }
\titlerunning{Jets of AGNs as Giant Coaxial Cables}

\maketitle

\section{Introduction}

Active galactic nuclei (AGNs) release vast amounts of energy, whose 
ultimate
source is a supermassive black hole in the galactic nucleus.  In so-called
radio-loud AGNs, two relativistic jets of plasma emanate from the nucleus,
presumably along the rotational axis of the black hole.  The radio emission
is synchrotron radiation, and can be linearly polarized up to about 75\% in
optically thin regions with uniform magnetic fields, with the polarization
angle $\chi$ orthogonal to the projection of the magnetic field {\bf B}
onto the plane of the sky (Pacholczyk 1970).

Very Long Baseline Interferometry (VLBI) yields radio images with very high
resolution, corresponding to linear sizes of the order of a parsec
at the typical distances of AGNs.  A structure with a compact ``core'' at
one end and a jet extending away from it predominates for radio-loud AGNs.
The VLBI jets are
virtually always one-sided, due to the relativistic aberration of the 
radiation in
the forward direction of the jets' motion: one jet approaches the Earth
and is highly boosted, while the receding jet is highly de-boosted.

A theoretical picture of the basic nature of this core--jet structure was
proposed by Blandford \& K\"onigl (1979), in which the ``core'' observed with
VLBI corresponds to the ``photosphere'' of the jet, where the
optical depth is near unity, $\tau \approx 1$, and the jet material makes
a transition from optically thick to optically thin. Although the orientation
of the observed polarization angle rotates $90^{\circ}$ to become parallel
to the synchrotron {\bf B} field in sufficiently optically thick regions,
this transition does not occur until an optical depth of $\tau \approx 6$,
so that the polarized emission observed in all regions, including the 
VLBI core, is effectively expected to be optically thin (Cobb 1993, 
Wardle 2018).

Multi-frequency VLBI polarization observations provide  information about
the wavelength dependence of the parsec-scale polarization. One example is
Faraday rotation occurring along the line of sight between the emitting
region and the observer.  Faraday
rotation is a rotation of the observed linear polarization that
arises when the associated electromagnetic wave passes through a region
with free electrons and a non-zero {\bf B} field component along the
line of sight. The simplest case corresponds to the
situation when this mechanism operates in regions of ``thermal''
(non-relativistic or only mildly relativistic) plasma outside
the emitting region, when the rotation is given by
\begin{eqnarray}
           \chi_{obs} - \chi_o =
\frac{e^3\lambda^{2}}{8\pi^2\epsilon_om^2c^3}\int n_{e}
{\mathbf B}\cdot d{\mathbf l} \equiv RM\lambda^{2}
,\end{eqnarray}
where $\chi_{obs}$ and $\chi_o$ are the observed and intrinsic
polarization angles, respectively, $-e$ and $m$ are the charge and
mass of the particles giving rise to the Faraday rotation, usually
taken to be electrons, $c$ is the speed of light, $n_{e}$ is the
density of the Faraday-rotating electrons, $\mathbf{B}$ is the magnetic
field, $d\mathbf{l}$ is an element along the line of sight, $\lambda$ is
the observing wavelength, $\epsilon_o$ is the permittivity of free space,
and the coefficient of $\lambda^2$ is called the Rotation Measure, RM (e.g.,
Burn 1966).
The action of such ``external'' Faraday rotation
can be identified using simultaneous multifrequency observations, through
its linear $\lambda^2$ dependence, allowing the determination of both the
RM (which reflects the electron density and line-of-sight {\bf B} field
in the region of Faraday rotation) and $\chi_o$ (the intrinsic direction
of the source's linear polarization, and hence the synchrotron {\bf B} field,
projected onto the plane of the sky).

Many theoretical studies and simulations of the relativistic jets of AGNs
have predicted the development of a helical jet {\bf B} field, which
essentially comes
about due to the combination of the rotation of the central black hole and
its accretion disk and the jet outflow; Tchekovskoy and Bromberg (2016)
provide a recent example.
Researchers have long been aware that the presence of a helical
jet {\bf B} field could give rise to a regular gradient in the observed
RM across the jet, due to the systematic change in the line-of-sight
component of the helical field (Perley et al. 1984, Blandford 1993).
Statistically significant transverse RM gradients across the parsec-scale
jets of an increasing number of AGNs have been reported in the literature over the past several years (Gabuzda et al. 2014, 2015b, 2017),
and have been interpreted as reflecting the systematic change in the line-of-sight
component of a toroidal or helical jet {\bf B} field across the jets 
(a helical {\bf B} field includes both toroidal and poloidal components;
it is the toroidal component that gives rise to the transverse RM gradient).
Basic physics leads to the conclusion that these jets carry currents,
whose direction can be inferred from the direction of the toroidal {\bf B}
field component giving rise to the transverse RM gradients.

Because the relativistic jets of AGNs are typically very narrow
structures, it is important to verify that RM structures across
jets that are not well resolved in the transverse direction are reliable.
As has been discussed earlier by Gabuzda et al. (2015b), the Monte Carlo 
simulations of Hovatta et al. (2012), Mahmud et al. (2013) and Murphy and
Gabuzda (2013)
have demonstrated that, for RM maps made at wavelengths in the range
2--6~cm, the probability that an individual RM gradient with a significance
of about $3\sigma$ or more is spurious is less than 1\%, even when the
observed width of the RM distribution across the jet is comparable to the
resolution (beam size).  These simulations also showed that RM gradients
with significances of $2-3\sigma$ can also be considered trustworthy
if they span at least two beam widths, or if they are observed at two or
more epochs.  In addition, simulated RM gradients remained visible even
when the intrinsic jet width was much less than the beam width (Mahmud et 
al. 2013; Murphy and Gabuzda 2013).

These results led to a series of analyses (Mahmud et al. 2013; Gabuzda et al.
2014, 2015b, 2017) focusing on (i) monotonicity of the RM gradients, 
(ii) the range of values encompassed by the gradients relative to the 
uncertainties in the RM measurements and (iii) steadiness of the change in 
the RM values across the jet (ensuring an apparent ``gradient'' is not due 
only to values in a few edge pixels). These were taken to be the key 
factors in determining the trustworthiness of
an RM gradient, that is, the probability that it is not spurious.  These were
all based on data obtained with the Very Long Baseline Array (VLBA)
at various sets of frequencies between 4.6~GHz and 15.3~GHz, and
include both reanalyses of data for previously published RM maps and the
publication of new RM maps. One key motivation for these studies was
the need for a meaningful statistical analysis of the directions of
the detected transverse RM gradients, that is, the directions
of the toroidal {\bf B}-field components, and thus of the directions of the
currents carried by AGN jets.

\begin{center}
\begin{table}[h!]
\begin{tabular}{c|c|c|}
\multicolumn{3}{c}{Table 1: Integrated RM values}\\
\hline
Source     &    Taylor et al. (2009) &  Hovatta et al. (2012)   \\
           &    Int. RM (rad/m$^2$)   & Int. RM (rad/m$^2$)   \\\hline
0059+581   &    $-175.4\pm 6.9$  & $-64.7$    \\
0133+476   &    $-101.2\pm 2.1$  & $-75.3$    \\
0212+735   &    $+21.8\pm 1.0$   & $+9.7$    \\
0403$-$132 &    $+7.1\pm 0.9$    & $+7.8$\\
0446+112   &    $+19.4\pm 14.4$ & $+21.9$\\
0552+398   &    $1.9\pm 5.5$    & $-0.6$ \\
0834$-$201 &    $-102.3\pm 8.7$ & $+45$  \\
0859$-$140 &    $17.0\pm 0.7$   & $+1.3$    \\
0955+476   &    $22.1\pm 4.0$   & $+9.6$\\
1124$-$186 &    $-10.2\pm 7.7$  & $-9.3$   \\
1150+812   &    $80.6\pm 3.2$   & $-19.4$  \\
1504$-$166 &    $-26.4\pm 10.1$ & $-11.2$  \\
1611+343   &    $14.9\pm 0.7$   & $+6.0$   \\
1641+399   &    $18.0\pm 0.3$   & $+19.4$   \\
1908$-$201 &    $-73.3\pm 1.8$  & $-20.4$  \\
2005+403   &    $-171.4\pm 11.5$& $-40.0$  \\
2216$-$038 &       *            &  $+0.6$    \\
2351+456   &    $-31.3\pm 5.0$  & $-33.5$ \\ \hline
\multicolumn{3}{l}{* No entry is given for this RM by Taylor et al. (2009;}\\
\multicolumn{3}{l}{the value given by Simard--Normandin et al. (1981) is}\\
\multicolumn{3}{l}{$-7\pm 9$~rad/m$^2$, consistent with the value of}\\
\multicolumn{3}{l}{Hovatta et al. (2012).}\\
\end{tabular}
\end{table}
\end{center}

\begin{center}
\begin{table}[t]
\begin{tabular}{c|c|c|c|c|c}
\multicolumn{6}{c}{Table 2: Map properties}\\
\hline
Source &  Peak  & Lowest & BMaj & BMin & BPA \\
     &    (Jy/beam) &  contour (\%) & (mas) & (mas) & (deg) \\\hline
0059+581 &  2.89  & 0125 & 1.40 & 1.20 & 49.8       \\
0059+581 &  2.88& 0125 & 1.30 & 1.30 &   0        \\\hline
0133+476 & 0.951 &  0.250& 1.39 & 1.17 & $-$4.7   \\
0133+476 & 0.954  & 0.250 & 1.28 & 1.28 &    0    \\ \hline
0212+735 & 3.01  & 0.250 & 1.28  & 1.22  & 43.8     \\
0212+735 & 3.01 & 0.250 & 1.25  & 1.25  &   0   \\\hline
0403$-$132 & 1.38     & 0.250&2.49   & 1.03  & 0.5    \\
0403$-$132 & 1.37     & 0.250& 1.60  &  1.60 &   0     \\\hline
0446+112 &  1.61   & 0.250& 2.15  & 1.03  & $-$1.5  \\
0446+112 &  1.66   & 0.250& 1.49  & 1.49  &   0     \\\hline
0552+398 & 3.36    & 0.125& 1.56  & 1.13  & $-$3.0      \\
0552+398 & 3.45    & 0.125& 1.33  & 1.33  &   0   \\\hline
0834$-$201 & 2.78    & 0.250& 2.79  & 0.87  & $-$3.3  \\
0834$-$201 & 2.98    & 0.250& 1.56  & 1.56  &  0      \\\hline
0859$-$140 & 0.627 & 1.00 & 2.91  & 1.01  & $-$5.7  \\
0859$-$140 & 0.601&  1.00& 1.71  & 1.71  &   0     \\\hline
0955+476 & 1.43    & 0.250& 1.48  & 1.21  & 1.7    \\
0955+476 & 1.44    & 0.250& 1.34  & 1.34  &   0     \\ \hline
1124$-$186 & 1.26    & 0.250& 2.90  & 0.94  & $-$2.7 \\
1124$-$186 & 1.26    & 0.250& 1.65  & 1.65  &   0    \\ \hline
1150+812 & 1.36   & 0.250& 1.22  & 1.16  & 3.1   \\
1150+812 & 1.36    &0.250 & 1.19  &  1.19 & 0      \\ \hline
1504$-$166 & 0.981   & 0.250& 2.79  & 0.86  & $-$4.3 \\
1504$-$166 & 0.868   & 0.500& 1.55  & 1.55  &    0   \\ \hline
1611+343 &  3.58   & 0.500& 1.64  & 1.14  & 1.5    \\
1611+343 &  3.54   & 0.500& 1.37  & 1.37  &   0    \\ \hline
1641+399 &  1.99   & 0.500& 1.59  & 1.15  & 4.7    \\
1641+399 &  2.06   & 0.500& 1.35  & 1.35  &  0     \\ \hline
1908$-$201 & 2.80    & 0.250& 2.68  & 0.93  & 1.4    \\
1908$-$201 & 2.82    & 0.250& 1.58  & 1.58  &   0    \\ \hline
2005+403 &  0.843  & 0.500& 1.78  & 1.26  & 21.4   \\
2005+403 &  0.863  & 0.500& 1.50  & 1.50  &    0   \\ \hline
2216$-$038 & 1.852   & 0.250& 2.42  & 0.98  & $-$4.7 \\
2216$-$038 & 1.854   & 0.250& 1.54  & 1.54  &   0    \\ \hline
2351+456 &   0.978 & 0.500& 1.60  & 1.12  & 18.5   \\
2351+456 &   1.005 & 0.500& 1.34  & 1.34  &    0    \\ \hline
\end{tabular}
\end{table}
\end{center}

\begin{figure*}[!htb]
\begin{center}
\includegraphics[width=0.38\textwidth,angle=90]{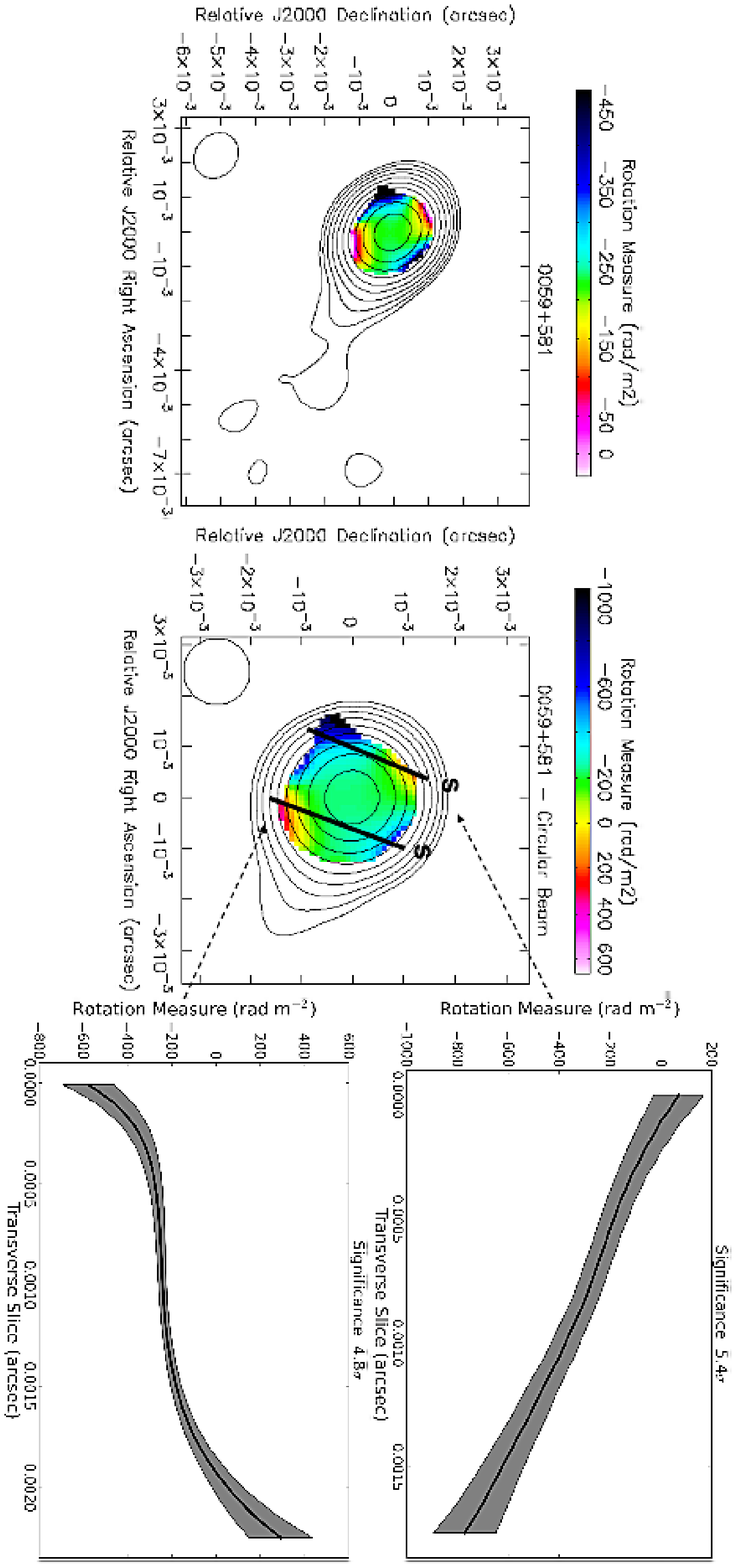}
\bigskip
\medskip
\includegraphics[width=0.38\textwidth,angle=90]{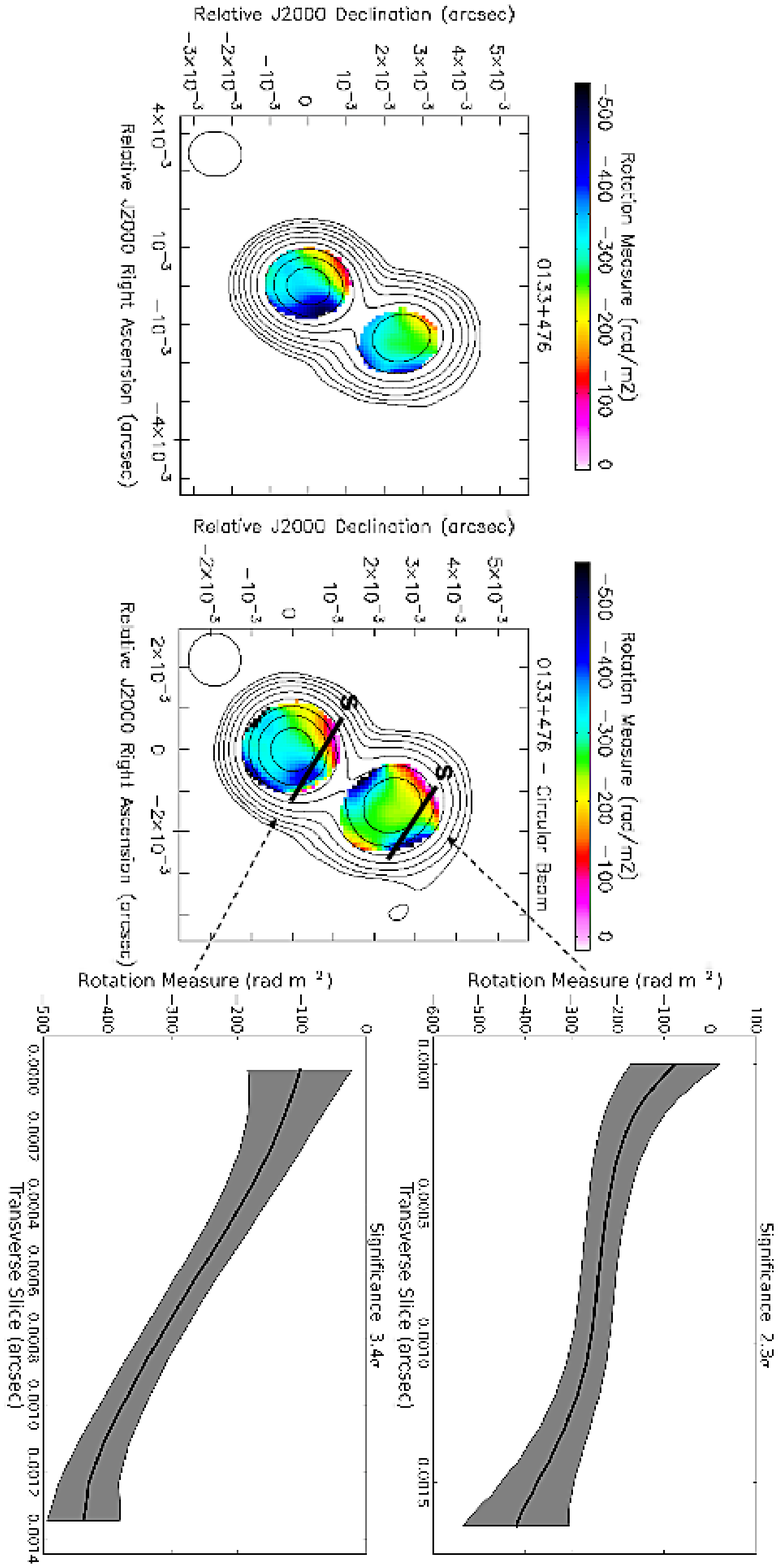}
\bigskip
\medskip
\includegraphics[width=0.38\textwidth,angle=90]{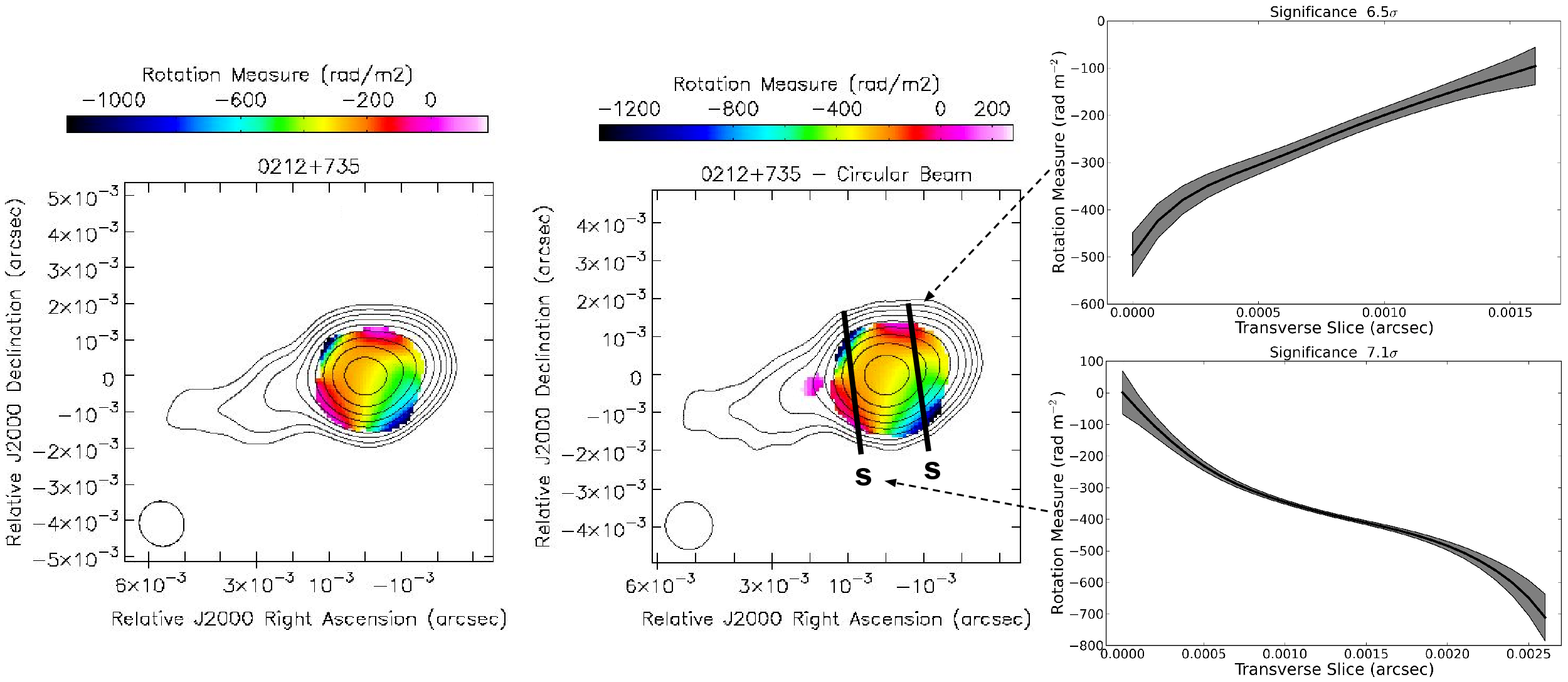}
\end{center}
\caption{8.1-GHz intensity maps made with the nominal, naturally weighted
elliptical beams with the corresponding RM distributions superposed (left),
the corresponding intensity and RM maps made using equal-area circular
beams (middle), and slices taken along the lines drawn across the RM
distributions in the middle panels (right).  These maps are based on the
8.1--15.2~GHz data of Hovatta et al. (2012); information about the map
peaks, bottom contours, and the beam sizes is given in Table~2.  Shown
here are results for 0059+581 (top), 0133+476 (middle) and 0212+735 (bottom). }
\label{fig:maps}
\end{figure*}

\begin{figure*}[!htb]
\begin{center}
\includegraphics[width=0.38\textwidth,angle=90]{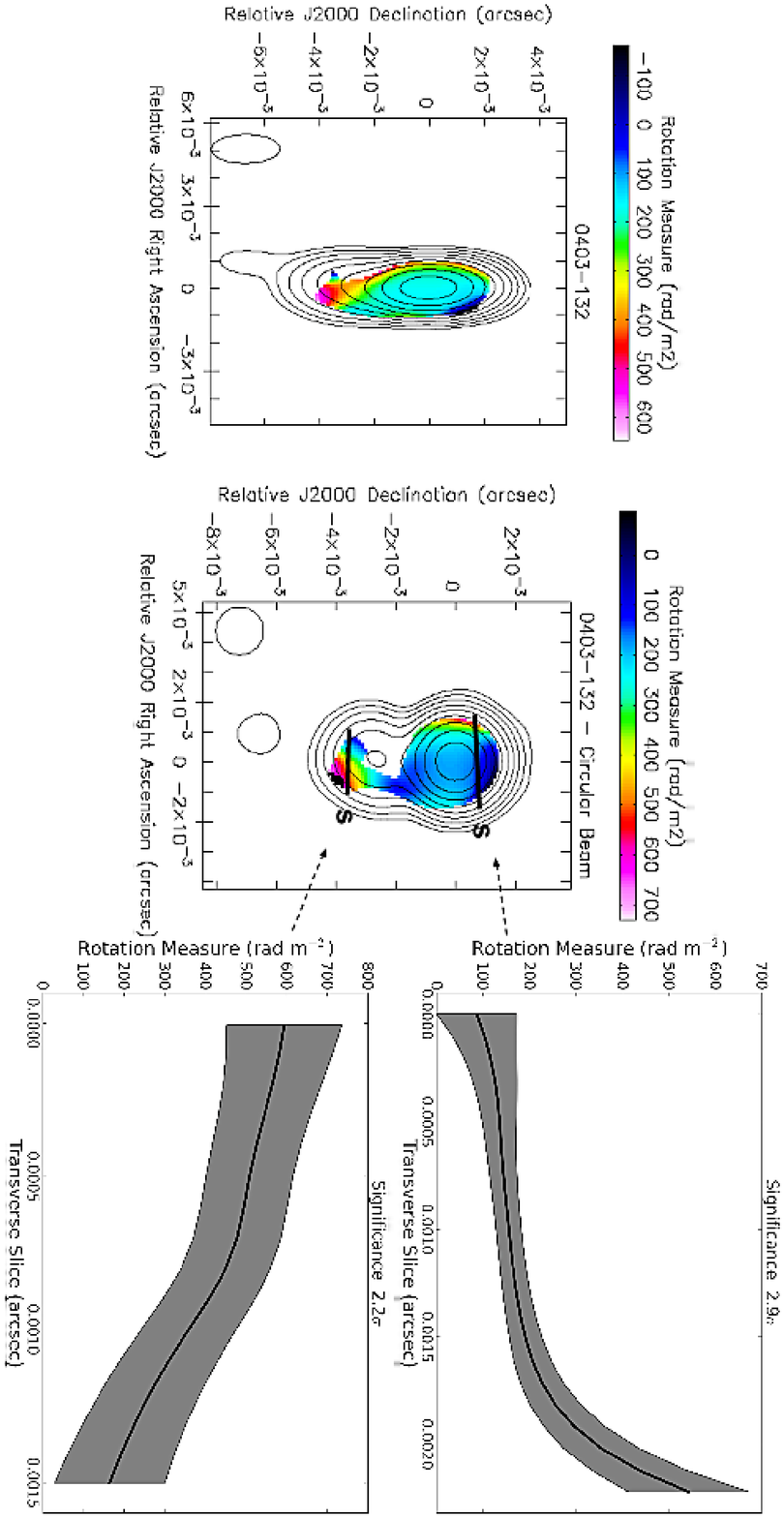}
\includegraphics[width=0.25\textwidth,angle=90]{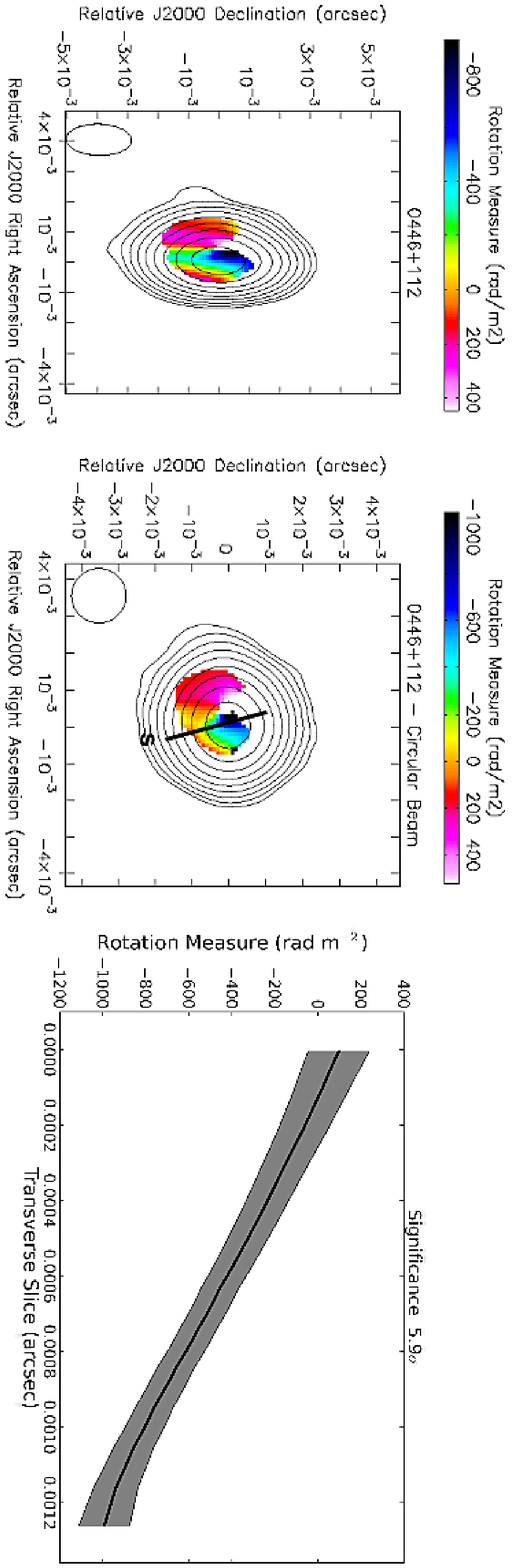}
\includegraphics[width=0.27\textwidth,angle=90]{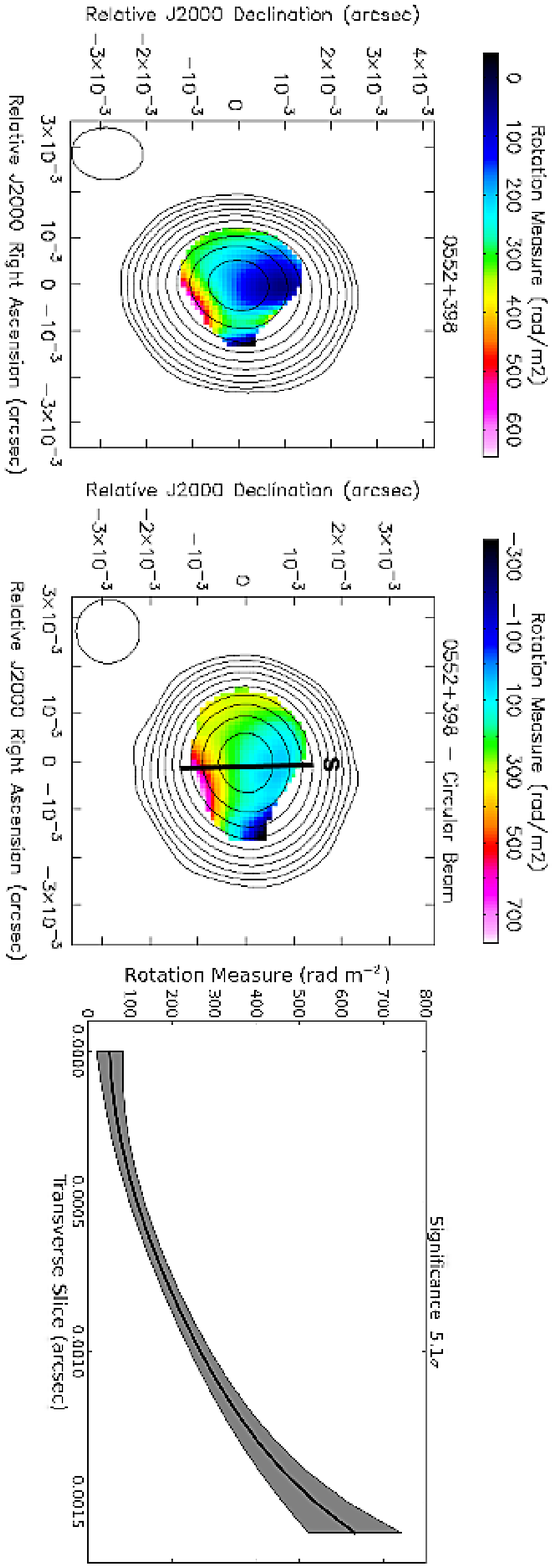}
\includegraphics[width=0.25\textwidth,angle=90]{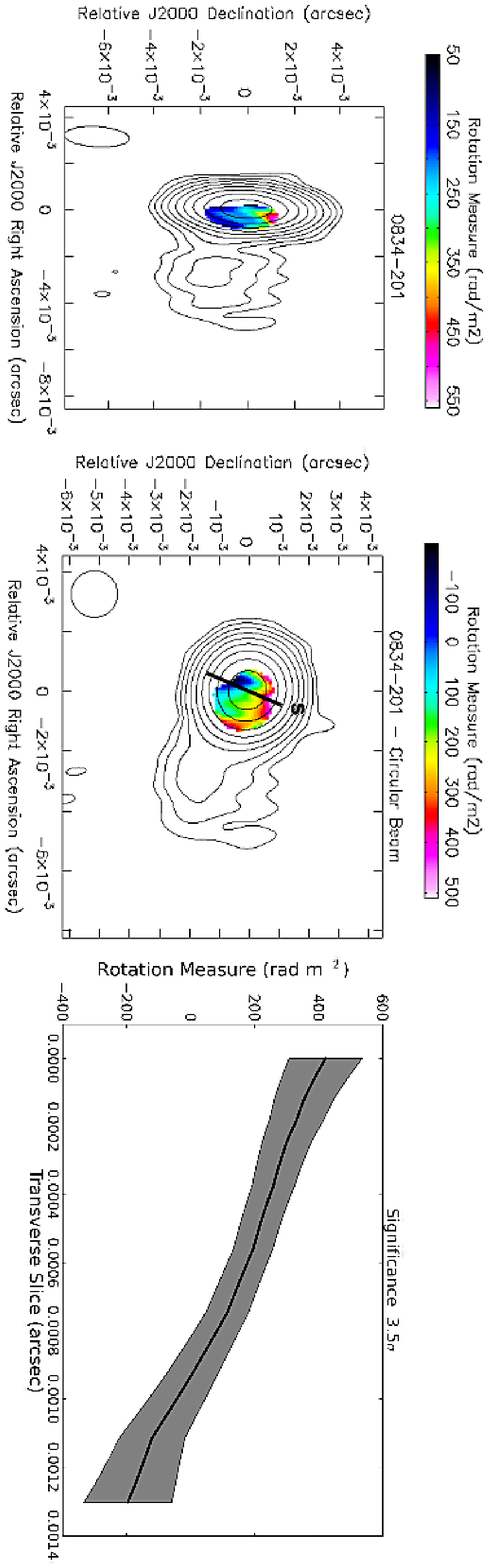}
\end{center}
\addtocounter{figure}{-1}
\caption{Continued. Results for 0403$-$132, 0446+112, 0552+398 and 0834$-$201
(from top to bottom).  }
\end{figure*}

\begin{figure*}[!htb]
\begin{center}
\includegraphics[width=0.38\textwidth,angle=90]{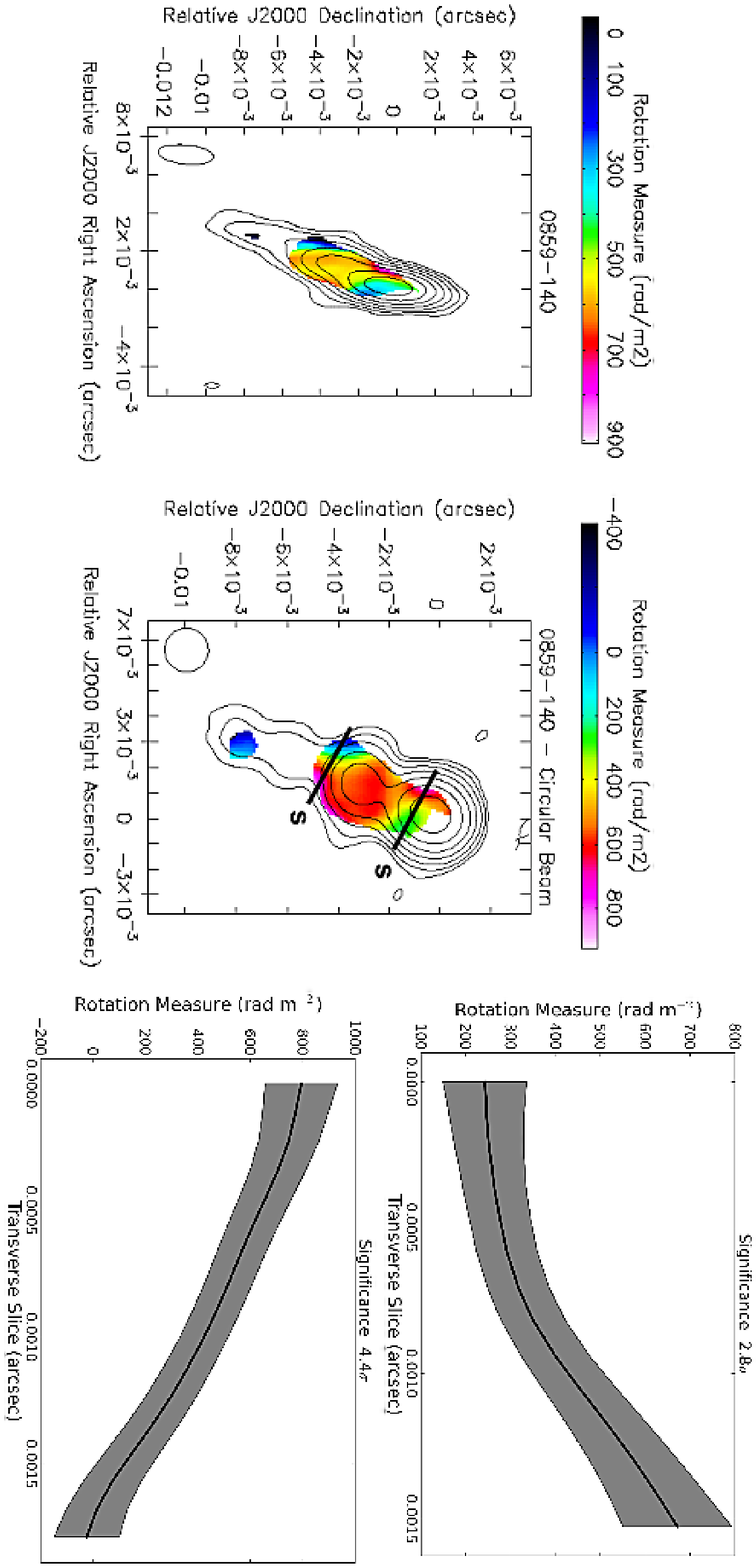}
\includegraphics[width=0.25\textwidth,angle=90]{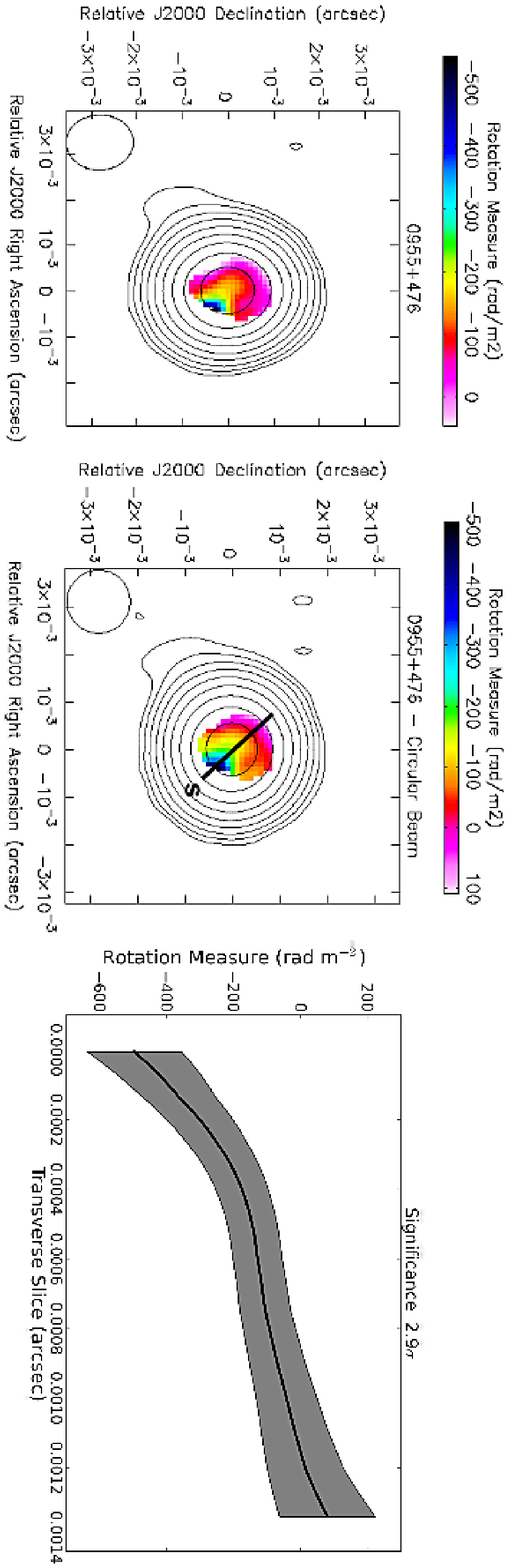}
\includegraphics[width=0.25\textwidth,angle=90]{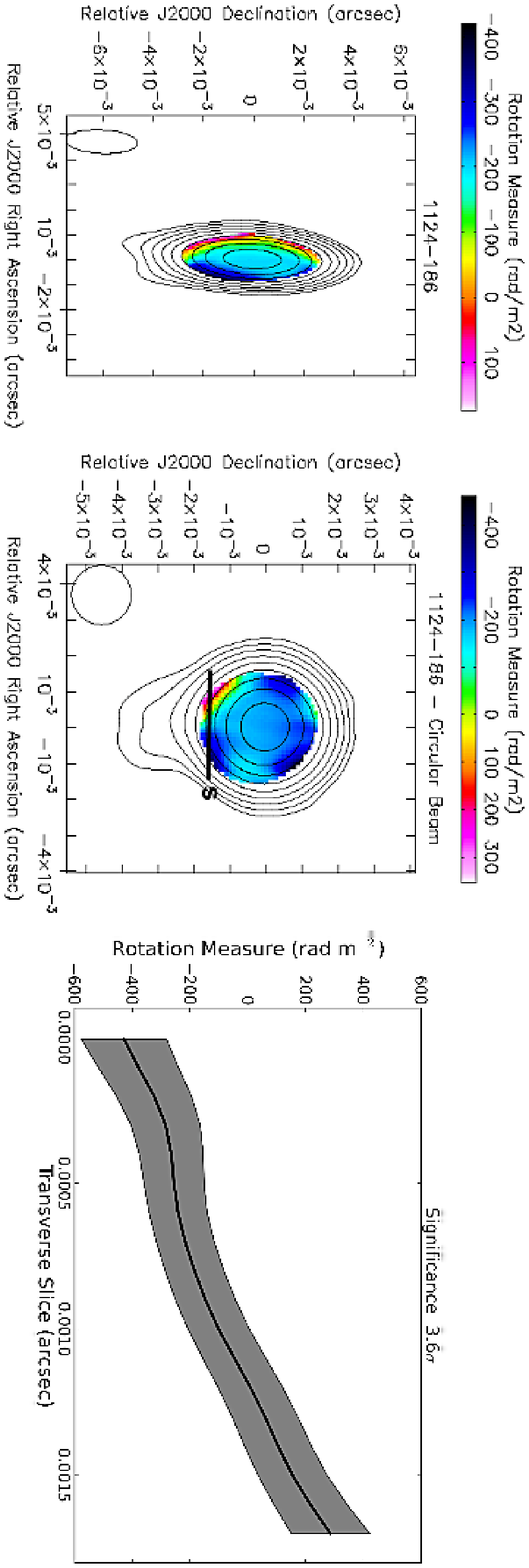}
\includegraphics[width=0.38\textwidth,angle=90]{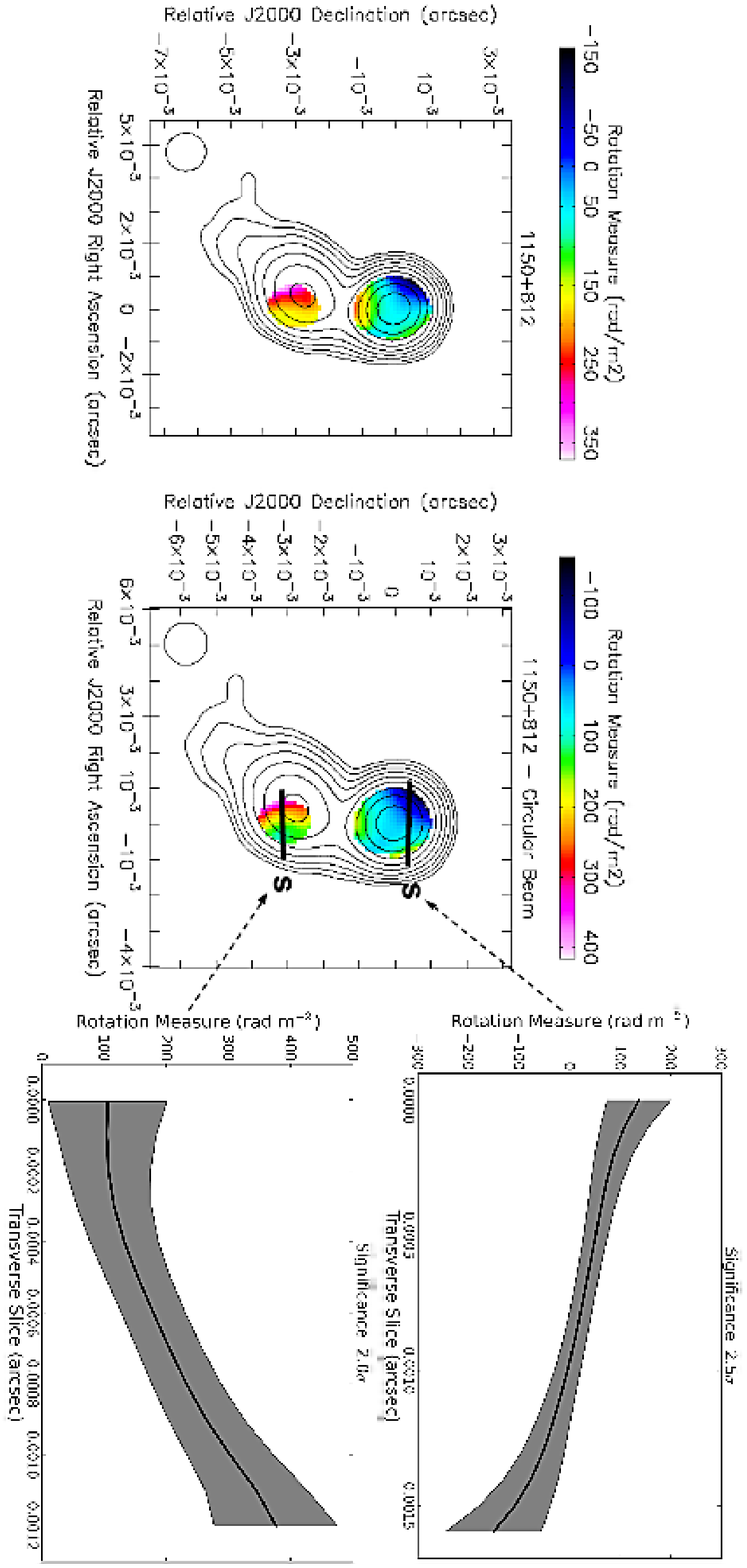}
\end{center}
\addtocounter{figure}{-1}
\caption{Continued. Results for 0859$-$140, 0955+476, 1124$-$186 and
1150+812 (from top to bottom).  }
\end{figure*}

\begin{figure*}[!htb]
\begin{center}
\includegraphics[width=0.25\textwidth,angle=90]{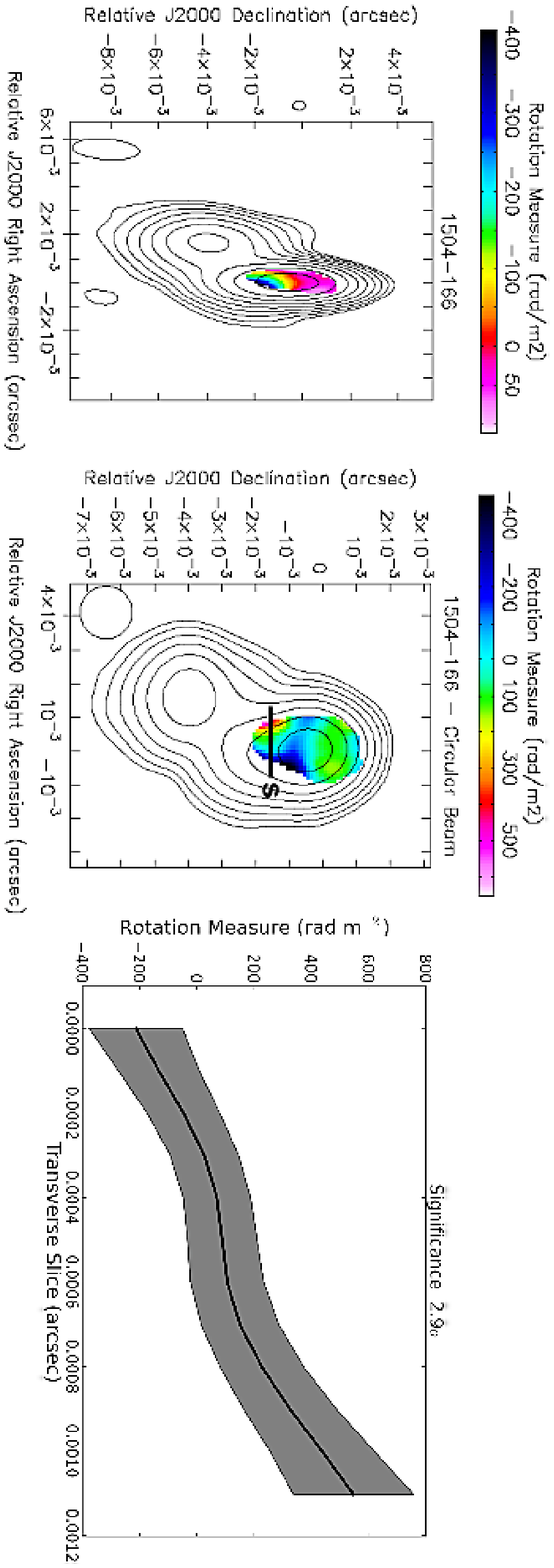}
\includegraphics[width=0.38\textwidth,angle=90]{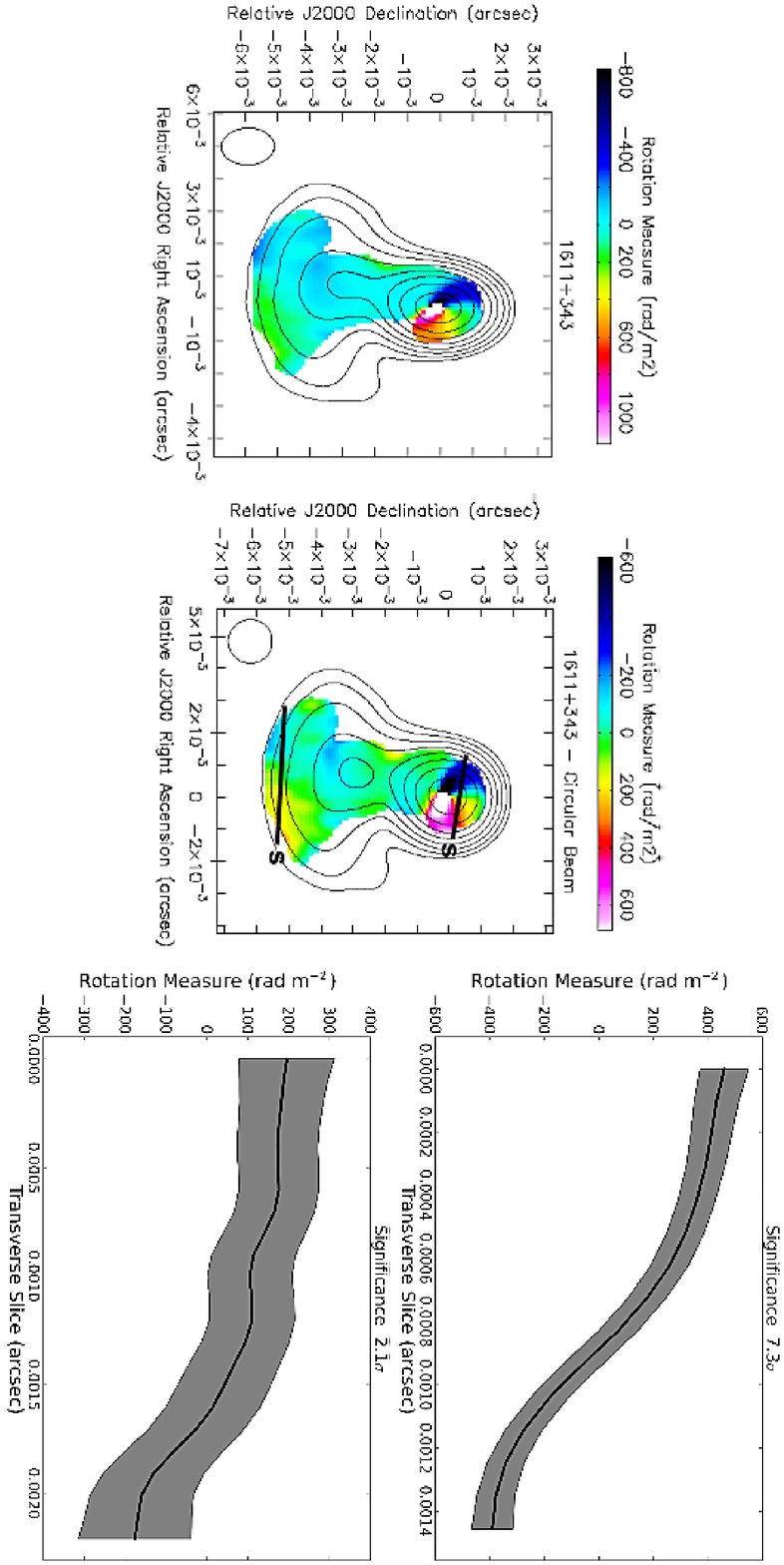}
\includegraphics[width=0.25\textwidth,angle=90]{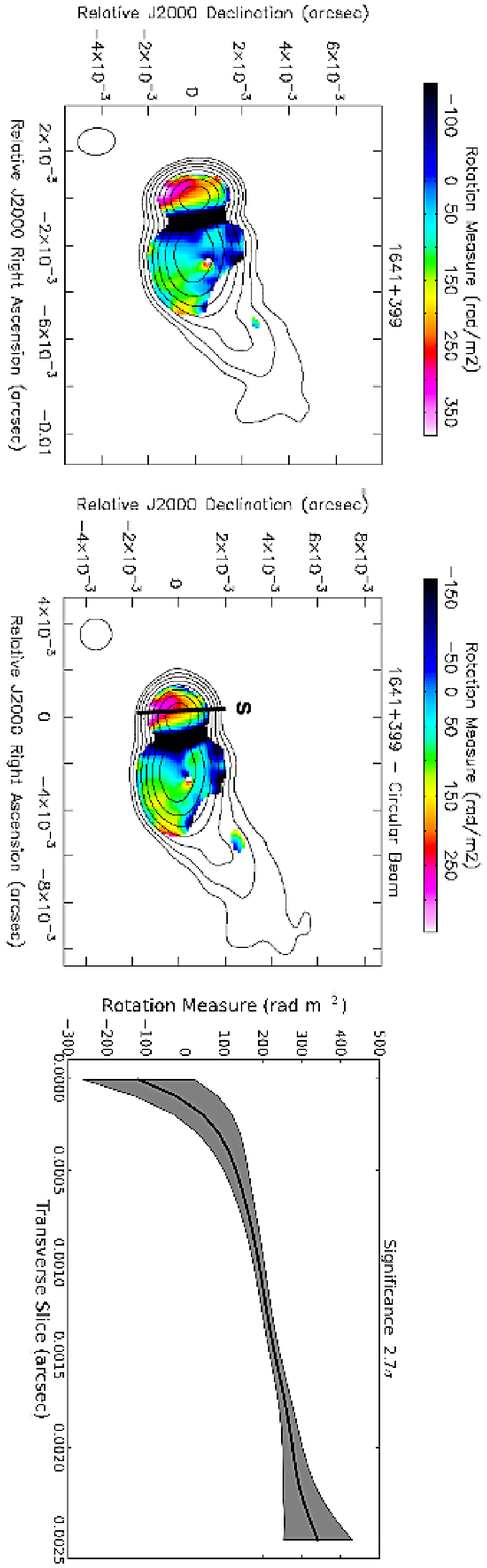}
\includegraphics[width=0.25\textwidth,angle=90]{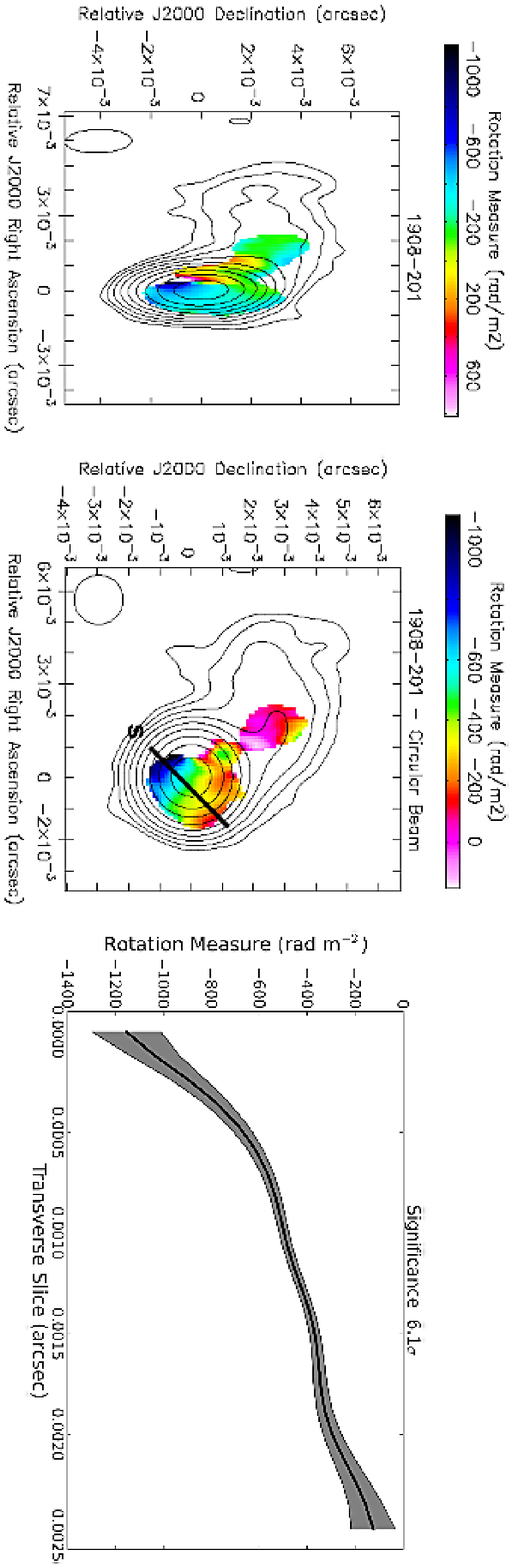}
\end{center}
\addtocounter{figure}{-1}
\caption{Continued. Results for 1504$-$166, 1611+343, 1641+399 and
1908$-$201 (from top to bottom).  }
\end{figure*}

\begin{figure*}[!htb]
\begin{center}
\includegraphics[width=0.38\textwidth,angle=90]{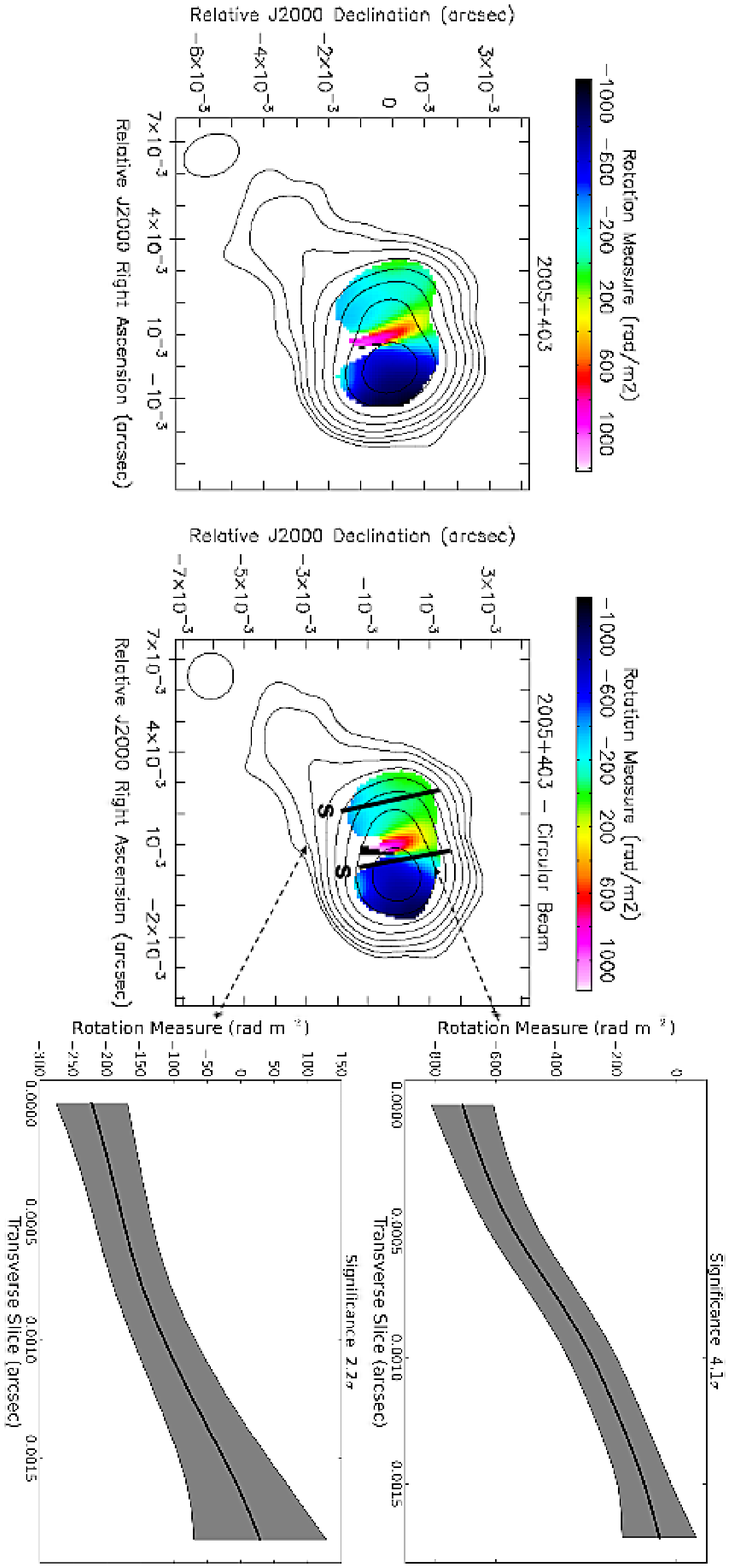}
\includegraphics[width=0.25\textwidth,angle=90]{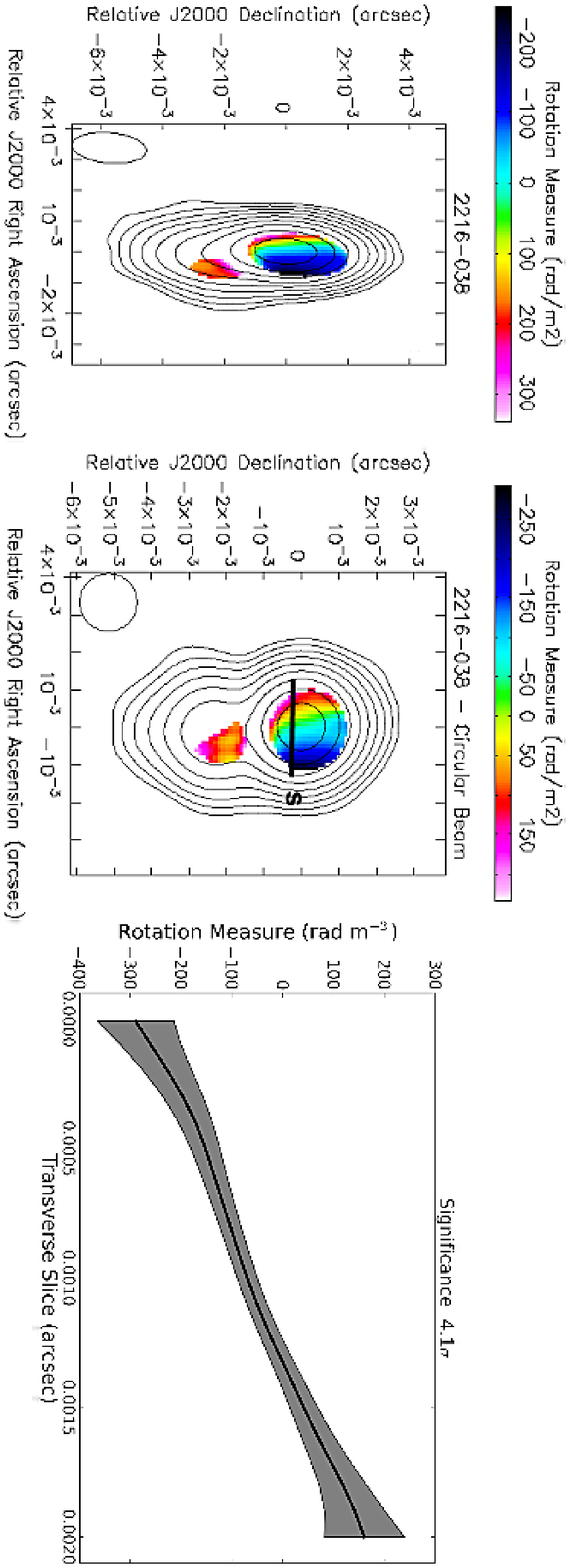}
\includegraphics[width=0.25\textwidth,angle=90]{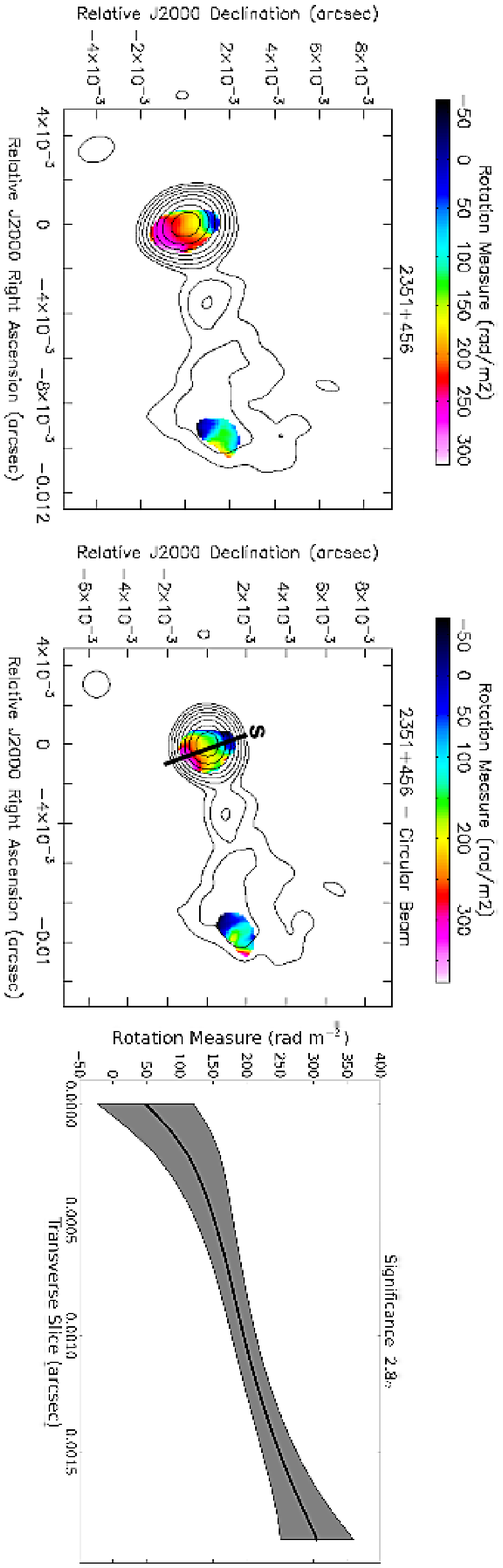}
\end{center}
\addtocounter{figure}{-1}
\caption{Continued. Results for 2005+403 (top), 2216$-$038 (middle)) and 
2351+456 (bottom).  }
\end{figure*}

\subsection{Outline of the paper}

Section~2 describes the observations and data reduction procedures used, in
particular how these are the same as and differ from those applied by
Hovatta et al. (2012). The results are presented in Section~3, and discussed
in Section~4. Our conclusions are summarized in Section~5.

\section{Observations and data reduction}

We carried out new analyses of the data used to produce the Faraday RM maps 
previously published by Hovatta et al. (2012).  These observations were 
obtained at 8.1, 8.4, 12.1 and 15.3~GHz in 2006 on the NRAO Very Long Baseline 
Array (VLBA). The self-calibrated visibility data were downloaded from the 
MOJAVE project website\footnote{http://www.physics.purdue.edu/MOJAVE/};
information about all steps of the calibration and initial imaging can be
found in Hovatta et al. (2012).

We carried out the following aspects of our analysis in the same way as 
Hovatta et al. (2012): 

{\bf Image alignment.} We aligned the polarization-angle images at the 
different frequencies
by determining the shifts between the corresponding intensity images
required to align their optically thin regions, in our case using a
cross-correlation method (Croke \& Gabuzda 2010).

{\bf Treatment of polarization-angle calibration errors.}
We did not include the systematic errors due to the absolute
polarization-angle calibration when searching for significant RM gradients,
since these cannot give rise to spurious RM gradients (Mahmud et al. 2009,
Hovatta et al. 2012).

{\bf Error estimation.}
We applied the improved error estimation formula of
Hovatta et al. (2012) to determine the uncertainties in the
polarization angles in individual pixels.

{\bf Treatment of Galactic Faraday Rotation.} 
We took into account the estimated Faraday rotation occurring in
our own Galaxy in the direction toward a source when significant, in order
to better estimate the RM distribution in the immediate vicnity of the source.

Our imaging and RM analysis differs from that of Hovatta et al. (2012)
in the following ways:

{\bf Technique used to match resolution.}
Hovatta et al. (2012) edited out the shortest baselines at the
higher frequencies and the longest baselines at the two lower frequencies
and used only the range of baselines common to all four datasets. In contrast,
because the range of frequencies analyzed is not large, we used all available
baselines, but ensured that the images at the different
frequencies were compared using a common beam (corresponding
to the lowest frequency).

{\bf Widths spanned by RM gradients.}
Hovatta et al. (2012) only searched for transverse RM gradients in RM
distributions spanning more than about 1.5 beamwidths across the VLBI jet.
Guided by the Monte Carlo simulation results described above (Hovatta et
al 2012, Mahmud et al. 2013, Murphy \& Gabuzda 2013),
we placed no
formal limit on the width spanned by an RM gradient (although in practice
only 4 of the 18 significant RM gradients we have detected are less than
1 beam width, and all are greater than 0.75 beam widths).

{\bf Treatment of core-region RM gradients.}
Hovatta et al. (2012) did not consider potential transverse RM gradients
in the vicinity of the VLBI core, on the grounds that this region was
at least partially optically thick, increasing the possibility of spurious
gradients. This issue has been discussed in a number of recent studies 
(Motter \& Gabuzda 2017, Gabuzda et al. 2017, Wardle 2018).
Because the expected $90^{\circ}$
``flip'' in the polarization angle in the transition from the optically
thin to the optically thick regime does not occur until an optical depth
of $\tau \simeq 6$, which is far upstream of the observed VLBI core region,
there is no reason to expect severe departures from a linear
$\lambda^2$ law for the Faraday rotation in the vast majority of cases,
and we, like a number of earlier studies (Gabuzda et al. 2014, 2015b, 2017;
Motter \& Gabuzda 2017), therefore analyzed
core-region RM gradients in the same way as
those located in the jet, having verified an absence of appreciable
deviations from a $\lambda^2$ law.

{\bf Estimation of significance.}
Hovatta et al. (2012) took the significance of an RM gradient to be
the absolute value of the difference in RM values at the two ends of the
gradient divided by the largest RM error at the edge of the jet (without
including the systematic error due to the absolute polarization-angle
calibration, which cannot give rise to spurious RM gradients, as was
noted above (Mahmud et al. 2009, Hovatta et al. 2012)). We used a somewhat 
more conservative approach to calculating the significance
of gradients, by comparing the absolute value of the RM difference across
a gradient with the square root of the sum of the errors on the two RM
values added in quadrature:

\begin{eqnarray}
\sigma & = & \frac{ |RM_1 - RM_2| }{ \sqrt{\sigma_1^2 + \sigma_2^2 }}
.\end{eqnarray}

Roughly speaking, our significances will usually be lower than those of
Hovatta et al. (2012), by up to about a factor of $\sqrt{2}$.
We note that in both of these approaches, the RM values considered will
be near the edges of the RM maps, where the uncertainties are highest,
so that both approaches are inherently conservative.

{\bf Supplementary analysis using circular beams.}
We considered versions of all the RM maps convolved using
circular beams with areas equal to those of the natural-weighted elliptical
beams (i.e., the radius of the circular beam was $R = \sqrt{(BMAJ)(BMIN)}$,
where $BMAJ$ and $BMIN$ are the major and minor axes of the elliptical
beam), as has been discussed in some earlier studies (Gabuzda et al.
2017, Motter \& Gabuzda 2017).
This helped test the robustness of gradients present in the RM maps made
using the elliptical beams, and in some cases helped clarify the direction
of a gradient relative to the local jet direction.

The observed Faraday rotation occurs predominantly in two locations: in the
immediate vicinity of the AGN and in our own Galaxy. The
latter ``Galactic'' RM must be estimated and removed if we
wish to identify Faraday rotation occurring in the vicinity of the
AGN itself.  Low-resolution, low-frequency integrated RM measurements will
generally be dominated by this Galactic RM component, which is uniform across
the source on milliarcsecond scales, and it is usually supposed that such
low-frequency, low-resolution (compared to the VLBI images) integrated RM
measurements provide a good estimate of the Galactic RM, although they will
formally include some contribution from plasma in the vicinity of the
source as well. Consistent with the approach adopted in the previous studies
considered here (Gabuzda et al. 2014, 2015b, 2017), we used the integrated RM
measurements of Taylor et al. (2009) at the positions of the sources
considered directly; these values, given in Table~1, are based on the 
VLA Sky Survey (NVSS) observations at two bands
near 1.4~GHz.  The integrated RMs for most of these AGNs are low, less than
about 30~rad/m$^2$ in absolute value; this is smaller than the typical
uncertainties in the RM values measured in our maps, and we did not
remove the effect of these small integrated RM values.  Six of the objects
have higher integrated RMs, ranging in absolute value from 80~rad/m$^2$
to 175~rad/m$^2$, and we removed these integrated RMs from all values in
the RM maps for these two sources. We note that this procedure is slightly
different to the approach of Hovatta et al. (2012), who removed the
average RM from the vicinity of the source on the sky based on the overall
RM image of Taylor et al. (2009), rather than the value at the position of
the source itself. This leads to small differences in the ranges of our RM
maps for these six sources compared to the previously published RM maps 
(Hovatta et al. 2012), although the maps are very similar overall. The 
Galactic RM
values estimated by Hovatta et al. (2012) and subtracted from their own
initial RM maps are also given in Table~1.

We tested the derived shifts between images at the different frequencies
by making spectral-index maps after applying  these shifts in order to
verify that they did not show any spurious features that could be due
to residual misalignment.  We also checked for broad consistency with the
shifts derived earlier for these same data (Pushkarev et al. 2012). Hovatta 
et al. (2012) also noted that their analyses confirmed that the appearance
of the spectral-index maps provides a good test of the correctness of the
relative alignment of the images at the various frequencies.

Hovatta et al. (2012) made RM maps for a sample of 191 extragalactic 
radio sources, but due to the limits they imposed on their analysis of 
possible transverse RM gradients, they only were able to identify transverse 
RM gradients that satisfied their criteria in  four sources. 
We visually inspected each of their 191 maps and identified for analysis
38 sources that showed possible transverse RM gradients by eye. For
each of these, we reconstructed the RM maps using the methods described
above and took RM slices in any region potentially displaying transverse
RM gradients, applying the new error estimation approach of Hovatta et al.
(2012) to estimate the RM uncertainties in individual pixels as input
to our estimate of the significances of any RM gradients analyzed.

We detected statistically significant, monotonic transverse RM gradients 
across the jets of 18 of the 38 sources investigated, which are considered
further below. The main factors contributing to the larger number of
such transverse RM gradients we were able to detect, compared to the much
smaller number reported by Hovatta et al. (2012), was the exclusion by
Hovatta et al. (2012) of transverse RM gradients arising within one beam width 
of the core, and spanning less than 1.5 beam widths across the jet.

\section{Results}

This paper represents the culmination of a series of studies carried out
by Gabuzda et al. (2014, 2015a, 2017). Here, we present the results of a 
reanalysis of data used to produce the RM images previously published
by Hovatta et al. (2012), based on VLBA data at 8.1, 8.4, 12.1 and 15.3~GHz,
together with an overall statistical analysis for the collected results
for transverse RM gradients across parsec-scale RM jets.

Our analysis of the 38 candidate AGNs with transverse RM gradients from
among the overall sample of 191 sources from the study of Hovatta et al.
(2012) indicated that 18 of these AGNs have firm, monotonic transverse RM
gradients (significances of
$2.8\sigma$ or greater, corresponding to a probability of occurring by
chance of no more than 0.5\%). Figure~1 presents 8.1-GHz intensity maps 
made with the nominal, naturally weighted
elliptical beams with the corresponding RM distributions superposed (left),
the corresponding intensity and RM maps made using equal-area circular
beams (middle), and slices taken along the lines drawn across the RM
distributions in the middle panels (right).  These maps are all based on the
8.1--15.2~GHz data of Hovatta et al. (2012); information about the map
peaks and bottom contours and the beam sizes is given in Table~2.

All 18 of these AGNs display statistically
significant transverse RM gradients across their jets, consistent with
these jet {\bf B} fields having significant toroidal components, possibly
associated with helical jet {\bf B} fields. Transverse RM gradients
are detected in 13 of these AGNs for the first time, while the remaining 5
cases correspond to AGNs in which transverse RM gradients have been observed
previously.

We note that, in large source
samples, a small number of spurious 3$\sigma$ gradients can appear purely
by chance. In a sample of 191 sources, with the probability that a 3$\sigma$
gradient is spurious being no larger than about 1\%, we would expect no more
than one or two such cases. Thus, it is possible that one or two of the 18
transverse RM gradients we have identified are spurious; thus,
the vast majority of the statistically significant transverse RM gradients we
have detected in these data represent real physical gradients, and the number
of possible spurious gradients is too small to affect our overall results.

\begin{center}
\begin{table}
\begin{tabular}{c|c|c|c}
\hline
\multicolumn{4}{c}{Table 3: Transverse RM Gradients}\\
Source &  Ref  &  Implied Current$^{\dagger}$ & Comments\\
       &       &                             & \\ \hline
\multicolumn{4}{c}{Firm, $\geq 2.8\sigma$ or confirmed}\\
0059+581 &    *       &   Out  & Reversal \\
0133+476 &    *       &   Out  & \\
0212+735 &    G17,*     &   In   & Reversal \\
0256+075 &    G15b     &   In   & \\
0300+470 &    G17     &   In   &\\
0305+039 &    G17     &   In   &\\
0333+321 &    G14     &   In   &\\
0355+508 &    G15b    &   Out  &\\
0403$-$132 &    *     &   In   & Possible reversal \\
0415+379 &    G17     &   In   &\\
0430+052 &   Go11&   In   &\\
0446+112 &    *       &   Out  &\\
0552+398 &    *       &   In   &\\
0716+714 &    M13       &   In   & Reversal \\
0735+178 &    G15b    &   In   &\\
0738+313 &    G14     &   Out  &\\
0745+241 &    G15b    &   In   &\\
0748+126 &    G15b    &   In   &\\
0820+225 &    G15b    &   In   &\\
0823+033 &    G15b    &   Out  &\\
0834-201 &    *       &   Out  &\\
0859-140 &    *       &   In   & Reversal \\
0923+392 &    G14     &   In  & Reversal \\
0945+408 &    G17     &   In  &\\
0955+476 &    *       &   In   &\\
1124-186 &    *       &   In   &\\
1156+295 &    G15b    &   In   &\\
1218+285 &    G15b    &   In   &\\
1226+023 &    H12       &   Out  &\\
1334$-$127&   G15b    &   Out  &\\
1502+106 &    G17     &   Out  &\\
1504-166 &    *       &   In   &\\
1611+343 &    G17,*     &   Out  & \\
1633+382 &    G14     &   In   &\\
1641+399 &    *,MG17  &   In   & \\
1652+398 &    G15b    &   Out  &\\
1749+096 &    G15b    &   In   &\\
1749+701 &    M13       &   In   & Reversal \\
1807+698 &    G15b    &   In   &\\
1908-201 &    *       &   In   &\\
2007+777 &    G15b    &   Out  &\\
2037+511 &    G14     &   In   & Reversal \\
2155$-$152&   G15b    &   In   &\\
2216-038 &    *       &   In   &\\
2230+114 &   H12        &   In   &\\
2345$-$167 &  G14     &   Out  &\\
2351+456 &    *       &   In   &\\
\multicolumn{4}{c}{Significant RM Gradients Showing Time Variability}\\
0836+710 &   G14,MG17   &  &\\
1150+812 &   G14,*       &  & Possible reversal\\
1803+784 &    M09      &  &\\
2005+403 &   G17,*          &  &\\
2200+420 &   G17,MG17     &     & \\ \hline
\multicolumn{4}{l}{G14 = Gabuzda et al. 2014; G15b = Gabuzda et al. 2015b;
G17 =}\\ 
\multicolumn{4}{l}{Gabuzda et al. 2017; Go11 = G\'omez et al. 2011; H12 = 
Hovatta}\\ 
\multicolumn{4}{l}{et al. 2012; M09 = Mahmud et al. 2009; M13 = Mahmud et 
al.}\\
\multicolumn{4}{l}{2013; MG17 = Motter \& Gabuzda 2017; * = this paper;}\\ 
\multicolumn{4}{l}{$\dagger$ Current corresponding to direction of innermost}\\ 
\multicolumn{4}{l}{gradient is given for sources displaying reversals.}\\
\end{tabular}
\end{table}
\end{center}

\section{Discussion}

A list of our 13 new and 5 confirmed transverse RM gradients together with 
all other known statistically
significant transverse RM gradients on parsec scales from the literature
(based on 5--15~GHz and 8--15~GHz RM maps) is given
in Table~3. We note that the other AGNs analyzed to obtain these collected
results are contained in subsamples of the 191 AGNs considered by
Hovatta et al. (2012), each containing no more than 40 AGNs, making it
unlikely that a significant number of the gradients detected in those samples
are spurious. Thus, we expect overall that at most two of the
transverse RM gradients listed in Table~1 are spurious.

The key importance of these new results is that they appreciably increase
the total number of AGNs in which firm transverse RM gradients have been found, 
raising this number to 52, enabling for the first time a reliable statistical
analysis of the directions of these transverse RM gradients on the sky; that is,
the directions of the jet currents implied by the toroidal {\bf B} field
components giving rise to the RM gradients.  Five of these 52 sources show
time variability in the directions of their transverse Faraday RM
gradients, and so cannot be used straightforwardly in such an analysis
(we will discuss a possible origin for this variability below).

Among the 47 sources for which the available data show no evidence of time
variability, the transverse RM gradients
in 33 imply inward jet currents and those in the remaining 14 outward
jet currents.  The probability of this asymmetry coming about
by chance can be estimated using a simple analysis based on a binomial
probability distribution: the probability of obtaining $N_{in}$
inward currents and $N_{out}$ outward currents in a total of $N_{in}+N_{out}$
jets by chance, if the probabilities of each of these jets having inward
or outward current are both equal to 0.5, is

\begin{equation}
P_{chance} =
\frac{1}{2^{(N_{in}+N_{out})}}\frac{(N_{in}+N_{out})!}{N_{in}!N_{out}!}.
\end{equation}

With $N_{in} = 33$ and $N_{out} = 14$, this yields a probability of
$0.243\%$. However, to accurately determine the total probability of such
an asymmetry in the number of inward and outward currents coming about
by chance, we must determine the total probability of 33 or more
of the 47 jets having inward current; summing all these individual probabilities
using the formula above yields a total probability of $0.40\%$.  This
low probability testifies that this asymmetry is a real physical
effect.

This asymmetry was in fact noted earlier by Contopoulos et al. (2009), but
their analysis was marred by a lack of
estimates of the significances for the 29 transverse RM gradients
considered, which were found from a purely visual inspection of RM maps in the
literature. This cast substantial doubt over those results. This has been
remedied by the analysis of the collected results presented here: in the
course of the analyses leading to the results listed in Table~3, the 
significances of all 29 potential RM gradients identified earlier by 
Contopoulos et al. (2009) by eye were checked:
of these 29 RM gradients, 22 proved to be statistically significant, 5
to be not statistically significant, and 2 to be statistically significant but
time variable. The fact that 24 of the
29 proved to be statistically significant simply reflects the fact that the
human eye is a reasonably good $3\sigma$ gradient detector. In addition to
clarifying which of the gradients considered earlier by Contopoulos et al.
(2009) are robust, the results in Table~3, including those presented in 
this paper, have added another 28 cases, more than doubling the available 
statistics.

We note that the absolute values of the observed RMs will be affected by 
the local Doppler factor, however, it is physically implausible for this to
give rise to systematic gradients in the observed RM across the jet. The 
Doppler factor depends on the intrinsic flow speed and the viewing angle,  
neither of which is expected to change systematically across the jet. An RM 
gradient could in principle be associated with a gradient in the electron 
density; however, electron-density gradients cannot give rise to the changes 
in the sign of the RM across the jets observed for a substantial number of the 
AGNs displaying transverse RM gradients. In addition, there is certainly no
physically plausible picture in which electron-density gradients could have
a preferred direction across the jet, whereas this comes about naturally if
there is a preferred direction for the jet current (i.e, for the jet 
toroidal {\bf B}-field component). 

Thus, it is now possible to assert with confidence that the
orientations of the toroidal {\bf B} field components of AGN jets on the sky
are not random, and that inward currents are considerably more
common in AGN jets on parsec scales than outward currents.

This result would be extremely interesting even on its own; however, it 
becomes truly intriguing when combined with the results of Christodoulou 
et al. (2016) and Knuettel et al. (2018), which indicate a strong, 
highly significantly significant predominance of {\em outward} currents 
implied by transverse RM gradients on decaparsec-to-kiloparsec scales: 
11 of 11 such gradients with significances $\geq 3\sigma$ correspond
to outward currents. The probability of this coming about by chance is only
$\simeq 0.05\%$.  
At first glance, it might seem
impossible to have predominantly inward jet currents on parsec scales,
but outward jet currents on larger scales.  However, here we must recall that
these currents can be distributed across the cross section of the
jet and surrounding space, and the effect that we are using to trace these
currents is the observed transverse RM gradients.

In fact, as was proposed earlier (Contopoulos et al. 2009, Mahmud et al.
2013) these collected results are consistent with a {\bf B}-field
configuration forming a nested helical-field structure, with one region of
helical field inside the other and with the two having oppositely directed
toroidal components. The orientation of the inner toroidal component
corresponds to inward currents along the jet, and that of the outer toroidal
component to outward currents along the jet direction, as shown
schematically in Fig.~2. This forms a system of currents and fields similar
to that of a co-axial cable, with inward current along the center of the
cable and outward current in a more extended sheath.
Both of the regions of helical field contribute to
the overall observed Faraday rotation; the inner region of helical field
makes the dominant contribution on parsec scales, while the outer region
of helical field makes the dominant contribution beyond a few tens of
parsec from the jet base. Changes in the conditions in different regions along
and across the jet with time could also explain the changes in the direction
of the significant transverse RM gradients observed in some sources as being due
to changes in whether the inner or outer region of helical {\bf B} field
dominates the overall Faraday-rotation integral at that location.

\begin{figure}
\begin{center}
\includegraphics[width=0.35\textwidth,angle=-90]{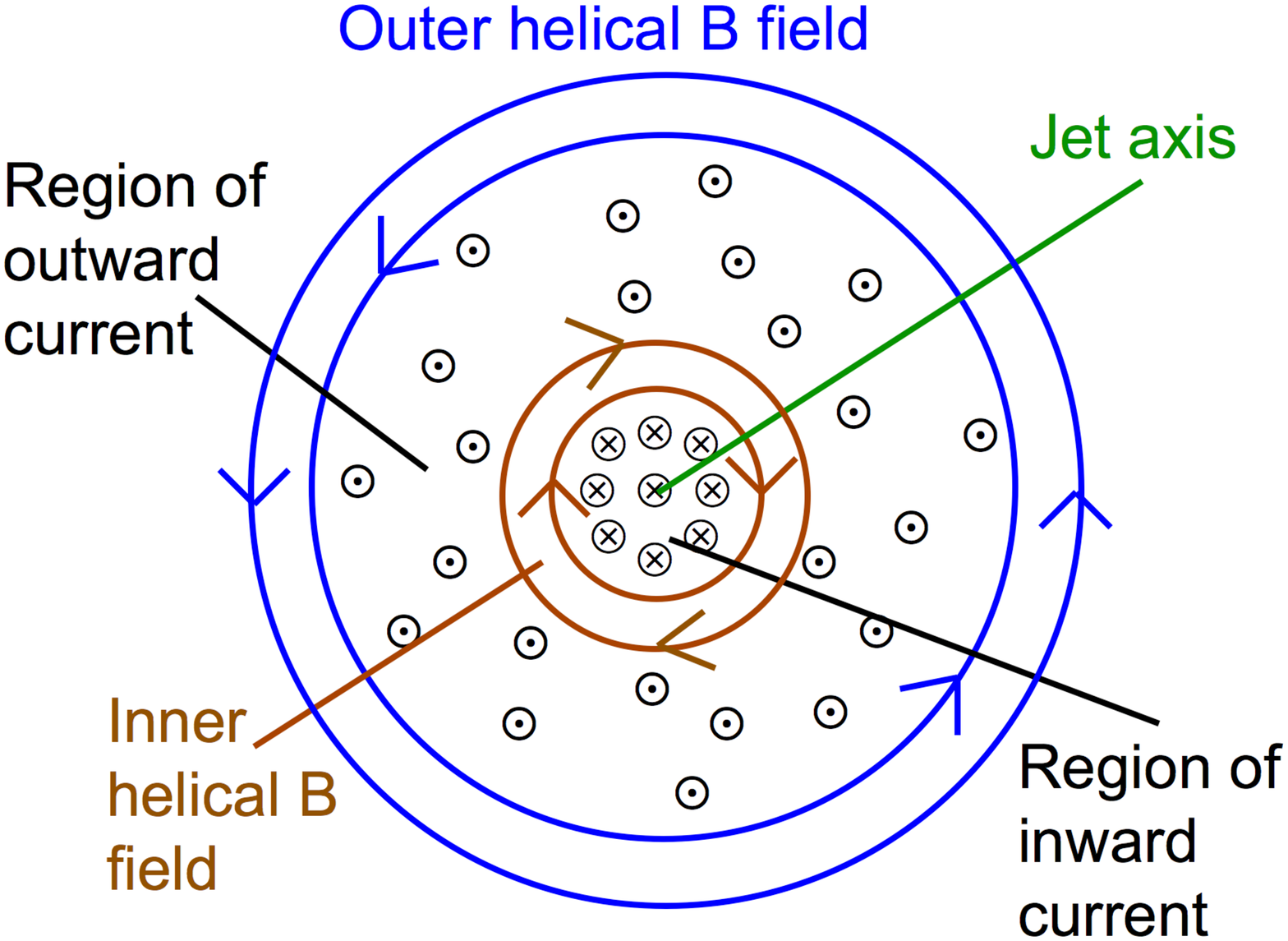}
\includegraphics[width=0.35\textwidth,angle=-90]{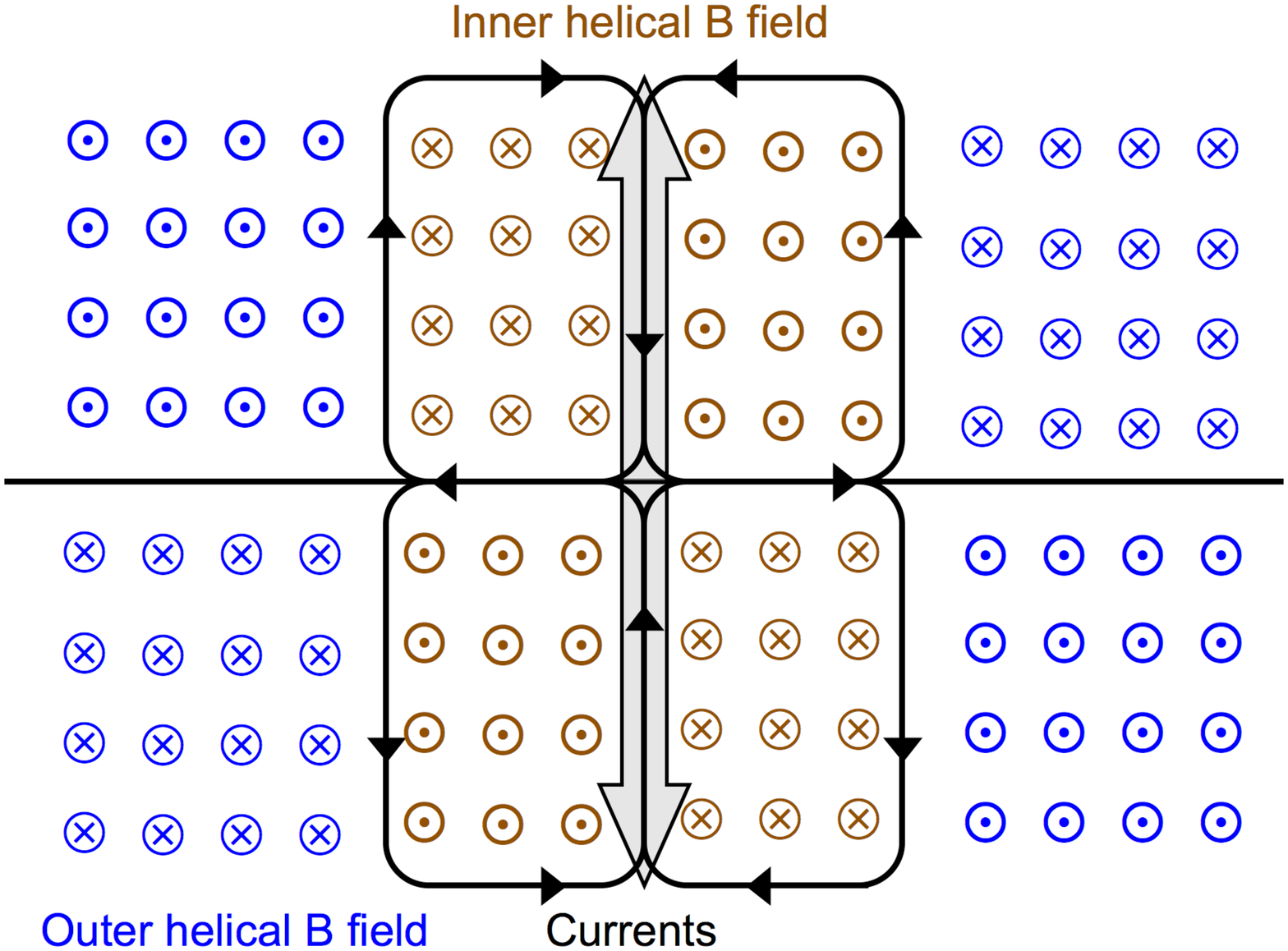}
\end{center}
\caption{Schematics of the system of {\bf B} fields and currents suggested
by the collected data on transverse Faraday rotation gradients, as viewed
from above (i.e., looking down the jet) (upper) and from the side (lower).
The region of inner helical field is shown in brown, the region of outer
helical field in blue, and the currents in black. The partially transparent
gray arrows in the lower panel represent the jet outflow. A circled dot
represents current or field oriented out of the page and a circled X
current or field oriented into the page.}
\label{fig:CB-fig}
\end{figure}

The inner and outer regions of helical field
correspond to the inner and outer sections of {\bf B} field loops that
have both been ``wound up'' by the differential rotation of the central
black hole and its accretion disk.  The distance at which each region 
dominates the observed transverse Faraday rotation gradients can be 
related to the scale on which the inner and outer (relative to the jet 
axis) sections of the initial loops of field are effectively ``wound up'' 
by the rotation (Christodoulou et al. 2016).
The current flows inward along the jet axis
and outward in a more extended region surrounding the jet, closing in the
kiloparsec-scale lobes and in the accretion disk.

The physical origin of this system of fields and currents is key to understanding the launching and propagation of astrophysical jets.
It is also important to note that the detection of statistically significant
transverse Faraday rotation gradients even out to kiloparsec scales 
(Gabuzda et al. 2015a, Christodoulou et al. 2016, Knuettel at al.  2018) 
demonstrates that an ordered helical or toroidal field component at least 
partially survives along
the entire length of the jet, though it may be masked by turbulence in
some regions.

It has already been pointed out (Contopoulos et al. 2009, Christodoulou
et al. 2016) that the configuration described above is in fact predicted by a
``cosmic battery'' model,
in which currents are generated in the accretion disk by the action of
the Poynting--Robertson effect, giving rise to a correlation between the
direction of the poloidal {\bf B} field that is ``wound up'' and the
direction of rotation of the accretion disk. This correlation leads to
a preferred orientation for the resulting toroidal {\bf B} field component,
which corresponds to an inward current along the jet axis. This mechanism
produces loops of magnetic field whose inner and outer parts
relative to the jet axis have opposite poloidal field directions, giving rise
to opposite toroidal field directions when they are wound up --- a nested
helical-field configuration such as has been proposed
as a way of explaining the opposite preferred
directions for the observed transverse RM gradients (toroidal {\bf B}-field
components) on parsec and decaparsec-kiloparsec scales.

\section{Conclusions}

We identify 13 new cases and confirm another 5 previously identified cases of statistically 
significant transverse
RM gradients across the parsec-scale jets of AGNs in the sample initially
analyzed by Hovatta et al. (2012). According to the Monte Carlo 
simulations of Hovatta et al. (2012) and Murphy \& Gabuzda (2013), the 
probability of a $\simeq 3\sigma$ RM gradient arising by chance in RM maps 
based on VLBA snapshot data at the frequencies considered here is no more 
than about 1\%; we accordingly expect no more than $\simeq 2$ of these 
gradients to be spurious, since the total sample contains 191 AGNs.

A statistical analysis of the collected results for the 47 AGNs that have
now been found to display statistically significant transverse RM
gradients on parsec scales indicates a clear preference for a
predominance of inward jet current on these scales. The probability that
this asymmetry has come about by chance is only $\simeq 0.40\%$.
In contrast, the collected results for AGNs displaying statistically
significant transverse RM gradients on larger (decaparsec-to-kiloparsec)
scales show a distinct predominance of outward currents, with an even
lower probability of this asymmetry arising by chance (Christodoulou et al.
2016, Knuettel et al. 2018). These results can be understood if the overall
systems of currents associated with AGN jets are similar to those of a
coaxial cable, with an inward current running down the center of the jet
and an outward current flowing in a more extended sheath-like region around 
the jet. This suggests the action of a ``battery'' mechanism that is
determining the direction of these currents (or equivalently the direction
of their associated toroidal {\bf B}-field components); one possibility
is the mechanism described by Contopoulos et al. (2009) and Christodoulou
et al. (2016).

A very interesting question that remains to be addressed is 
why a sizeable minority of the 47 AGNs listed in Table~2 have transverse
RM gradients on parsec scales whose directions imply outward rather
than inward currents. One possibility is that the transition from
dominance of the inner to dominance of the outer region of helical {\bf B} 
field expected in the cosmic-battery model occurs on smaller angular scales 
than those probed by the available observations in these sources. It is
also possible that the physical conditions in the accretion disks of some
AGNs are less conducive to efficient operation of the battery mechanism; in
this case, some fraction of the observed RM gradients would in fact have 
random directions, 
with an excess corresponding to inward current arising due to the operation 
of some mechanism that facilitates the development of inward jet currents.
It may also be that the direction of the jet current is also affected by other 
factors, such as Hall currents, as was suggested by K\"onigl (2010). Finally,
the numerical computations of Koutsantoniou and Contopoulos (2014) suggest
that the cosmic battery described above may operate in reverse when the 
central black hole rotates at more than about 70\% of its maximal rotation;
in such a case, the inner edge of the accretion disk would be close enough 
to the black hole horizon for the effect of the rotation of space--time to 
dominate over the effect of the disk's rotation, leading to a reversal in 
the direction of the radiation force on the electrons in the accretion disk 
in the toroidal direction. Future studies aimed at characterizing the 
properties of
the minority of AGNs whose transverse RM gradients imply outward currents
on parsec scales and possible differences from the properties of the majority 
of AGNs whose jets are observed to have inward currents are certainly of 
interest.

Whether or not the particular ``cosmic battery'' model described above is 
indeed operating in
the jet--accretion disk systems of AGNs has not been settled. However, our
results have now yielded firm evidence that many --- possibly all --- AGN
jets have inward currents along their axes and outward currents in a more
extended region surrounding the jets.  This provides fundamental information
about the conditions leading to the formation and launching of the jets,
as well as key input to theoretical simulations of astrophysical jets.
It also indicates that astrophysical jets are fundamentally electromagnetic
structures, which must be borne in mind when interpreting observed features
in the distributions of both their intensity and linear polarization.

\begin{acknowledgements}

We thank the referee for his/her clear and pertinent comments, which have
helped improve the overall clarity of this paper.
\end{acknowledgements}

\end{document}